\documentclass[3p,times,twocolumn]{elsarticle}
\biboptions{comma,compress}
 
\usepackage{graphicx}
\usepackage{ecrc}

\volume{}
\firstpage{1}
\jid{nppp}
\jnltitlelogo{Nuclear and Particle Physics Proceedings}
\journalname{Nuclear and Particle Physics Proceedings}
\runauth{T.\ Husek}

\usepackage[caption=false]{subfig}
\usepackage{amsmath}

\usepackage[unicode=true,colorlinks,breaklinks]{hyperref}
\hypersetup{pdftitle={Radiative corrections for Dalitz decays of \textbackslash pi\^{}0, \textbackslash eta\^{}(') and \textbackslash Sigma\^{}0}}

\makeatletter
\providecommand*{\diff}%
  {\@ifnextchar^{\DIfF}{\DIfF^{}}}
\def\DIfF^#1{%
  \mathop{\mathrm{\mathstrut d}}%
    \nolimits^{#1}\gobblespace}
\def\gobblespace{%
    \futurelet\diffarg\opspace}
\def\opspace{%
    \let\DiffSpace\!%
    \ifx\diffarg(%
      \let\DiffSpace\relax
     \else
      \ifx\diffarg[%
	\let\DiffSpace\relax
      \else
	\ifx\diffarg\{%
	  \let\DiffSpace\relax
	\fi\fi\fi\DiffSpace}

\begin{document}

\begin{frontmatter}
\title{Radiative corrections for Dalitz decays of $\pi^0$, $\eta^{(\prime)}$ and $\Sigma^0$}
\author[add]{Tom\'a\v{s} Husek}
\ead{thusek@ific.uv.es}
\address[add]{IFIC, Universitat de Val\`encia -- CSIC, Apt.\ Correus 22085, E-46071 Val\`encia, Spain}
\pagestyle{myheadings}
\markright{ }

\begin{abstract}
We briefly summarize current experimental and theoretical results on the $\pi^0$ Dalitz decay, including the new value for the ratio $R={\Gamma(\pi^0\to e^+e^-\gamma(\gamma))}/{\Gamma(\pi^0\to\gamma\gamma)}=11.978(6)\times10^{-3}$, which is by two orders of magnitude more precise than the current PDG average.
Furthermore, we discuss radiative corrections for the Dalitz decays $\eta^{(\prime)}\to\ell^+\ell^-\gamma$ beyond the soft-photon approximation.
The corrections inevitably depend on the $\eta^{(\prime)}\to\gamma^*\gamma^{(*)}$ transition form factors.
Finally, we present (inclusive) radiative corrections for the $\Sigma^0\to\Lambda e^+e^-$ decay, evaluated as well beyond the soft-photon approximation, i.e., over the whole range of the Dalitz plot and with no restrictions on the energy of the radiative photon.
Here, we also calculate explicitly the 1$\gamma$IR contribution and the correction to the $\Sigma^0\Lambda\gamma$ vertex and confirm that these can be neglected.
\end{abstract}

\end{frontmatter}

\section{The neutral-pion Dalitz decay}

The neutral-pion Dalitz decay $\pi^0\to e^+e^-\gamma$, the second most significant decay channel of $\pi^0$, allows us to access information about the pion transition form factor.
Moreover, it is used as the normalization channel in the rare-pion-decay ($\pi^0\to e^+e^-$) search, and its eventual inaccurate knowledge is a limiting factor for rare-kaon-decay measurements, such as $K^+\to\pi^+e^+e^-$ or $K^\pm\to\pi^\pm\pi^0e^+e^-$.
In general, to provide relevant experimental results, the correct and consistent incorporation of properly calculated radiative corrections is crucial.
Radiative corrections to the total decay rate of the $\pi^0$ Dalitz decay were first (numerically) addressed in Ref.~\cite{Joseph:1960zz}.
A pioneering study of the corrections to the differential decay rate was published by Lautrup and Smith~\cite{Lautrup:1971ew}, although only in the soft-photon approximation, and extended later in the work of Mikaelian and Smith~\cite{Mikaelian:1972yg}.

\begin{figure}[!t]
\centering
\subfloat[][]{
\includegraphics[width=0.4\columnwidth]{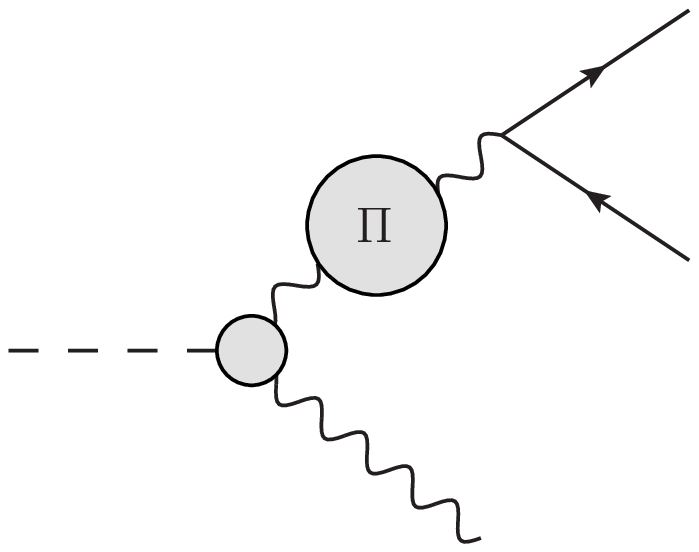}
\label{fig:virta}
}
\subfloat[][]{
\includegraphics[width=0.4\columnwidth]{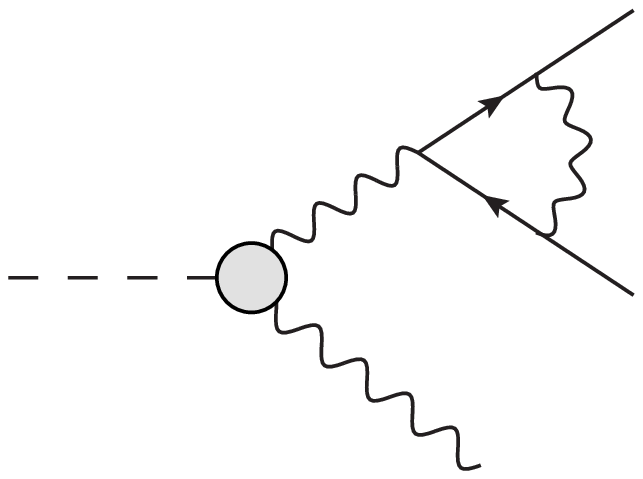}
\label{fig:virtb}
}

\subfloat[][]{
\includegraphics[width=0.4\columnwidth]{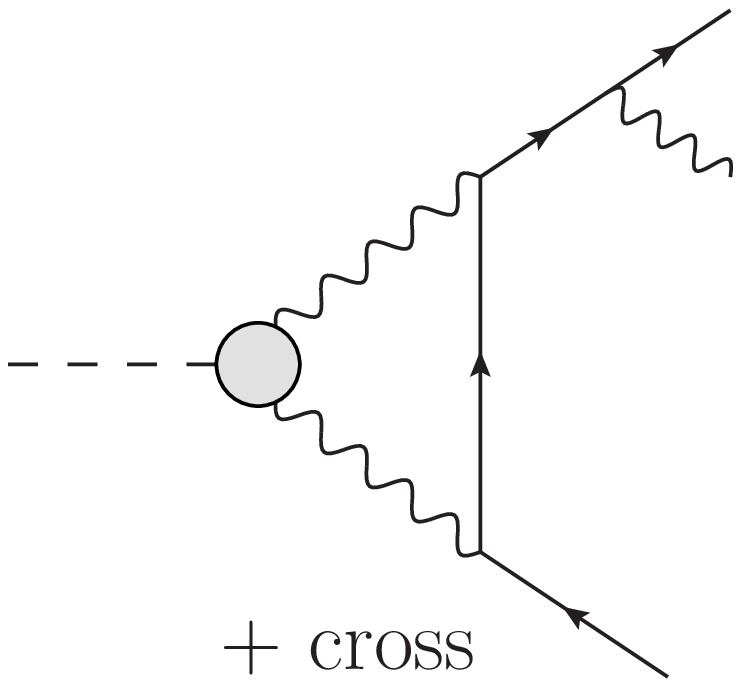}
\label{fig:1gIRa}
}
\subfloat[][]{
\includegraphics[width=0.4\columnwidth]{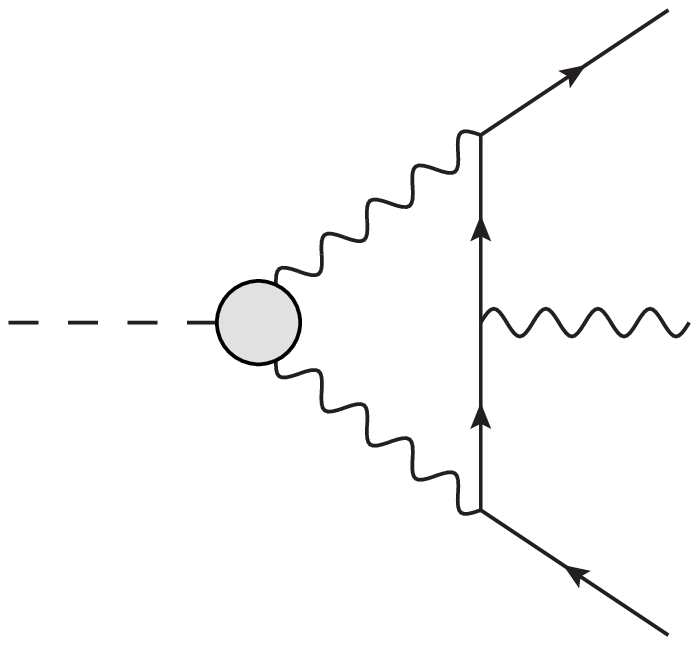}
\label{fig:1gIRb}
}

\subfloat[][]{
\includegraphics[width=0.9\columnwidth]{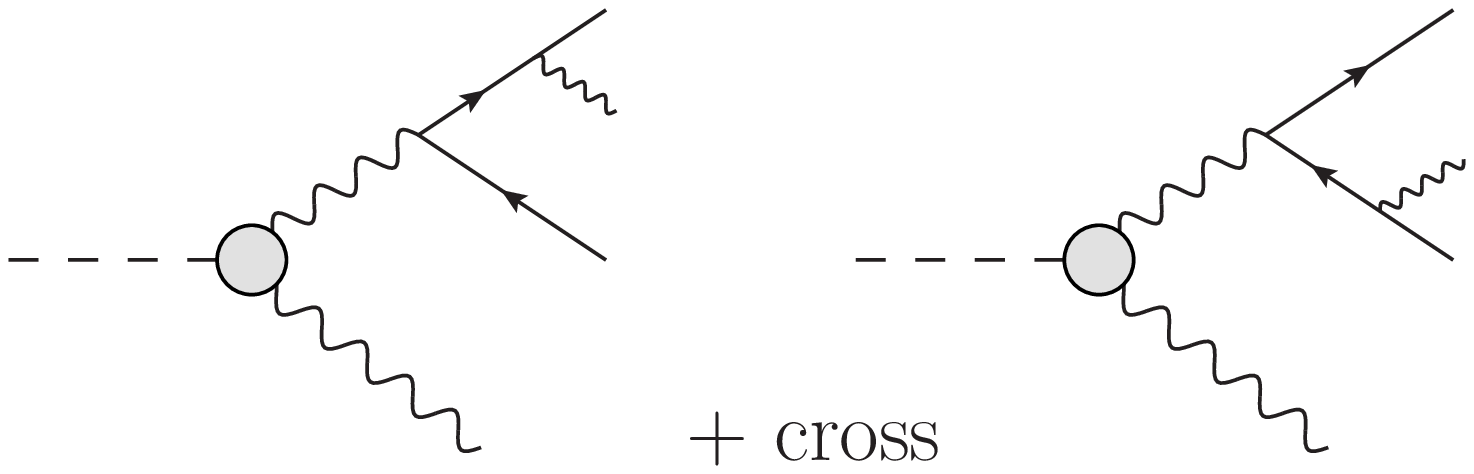}
\label{fig:BS}
}
\caption{
\label{fig:mesons}
NLO QED radiative corrections for the Dalitz decay $P\to \ell^+\ell^-\gamma$: a) vacuum polarization insertion, b) correction to the QED vertex, c) \& d) one-loop one-photon-irreducible contributions, e) bremsstrahlung.
Note that `cross' in figure (c) corresponds to a diagram where the photon is emitted from the outgoing positron line and `cross' in figure (e) stands for the diagrams with outgoing photons interchanged.
}
\end{figure}

A reinvestigation~\cite{Husek:2015sma} of this old topic was motivated by the needs of NA48/NA62 experiments at CERN.
Among other goals, the slope $a_\pi$ of the singly-virtual electromagnetic transition form factor $\mathcal{F}_{\pi^0\gamma^*\gamma^*}(0,q^2)$ was expected to be directly measured with unprecedented precision.
Compared to Ref.~\cite{Mikaelian:1972yg}, we took into account the one-photon-irreducible (1$\gamma$IR) contribution (see Fig.~\ref{fig:1gIRa} and \ref{fig:1gIRb}).
This contribution was, due to inappropriate assumptions and arguments based on the Low's theorem~\cite{Low:1958sn,Adler:1966gc,Pestieau:1967snm}, considered to be negligible.
However, when the exact calculation is performed, the 1$\gamma$IR contribution turns out to be significant~\cite{Husek:2015sma,Tupper:1983uw,Tupper:1986yk,Kampf:2005tz}.
Finally, we developed a \verb!C! code, which returns the radiative correction for any given kinematically allowed point and which propagated to the Monte Carlo event generator within the NA62 experiment.
Afterwards, 1.1$\times10^6$ fully reconstructed $\pi^0$ Dalitz decays were analyzed with the result $a_\pi^\text{NA62}=3.68(57)\,\%$~\cite{TheNA62:2016fhr}.
Let us note that the current PDG value $a_\pi^\text{PDG}=3.35(31)$~\cite{Tanabashi:2018oca} is dominated by two distinct types of inputs: The above NA62 precise time-like-region result and the value provided by the CELLO collaboration ($a_\pi^\text{CELLO}=3.26(37)\,\%$)~\cite{Behrend:1990sr} using the (model-dependent) extrapolation from the space-like region.

Having at hand the complete set of the next-to-leading-order (NLO) QED radiative corrections $\delta(x,y)$ (for diagrams see Fig.~\ref{fig:mesons}) and a relatively good knowledge of the form-factor shape (in the neutral-pion case represented by the form-factor slope $a_\pi$), we can determine very precisely (and at the same time reliably) the following ratio~\cite{Husek:2018qdx}:
\begin{equation}
\begin{split}
&R=\frac{\Gamma(\pi^0\to e^+e^-\gamma(\gamma))}{\Gamma(\pi^0\to\gamma\gamma)}
\simeq\frac{\alpha}{\pi}
\iint\diff x\diff y\left\{\vphantom{\frac12}\right.\\
&\left.(1+a_\pi x)^2
(1+\delta(x,y))
\frac{(1-x)^3}{4x}
\left[1+y^2+\frac{4m_e^2}{M_\pi^2x}\right]
\right\}.
\label{eq:R}
\end{split}
\end{equation}
Above, the kinematic variables $x$ and $y$ are defined in the following way: $x={(p_{e^-}+p_{e^+})^2}/{M_\pi^2}$ and $y=2{p_{\pi^0}\cdot(p_{e^+}-p_{e^-})}/[{M_\pi^2}{(1-x)}]$\,.
With a rather conservative choice $a_\pi^\text{univ}\equiv3.55(70)\,\%$ for the slope we find $R=11.978(5)(3)\times10^{-3}$~\cite{Husek:2018qdx}; the former uncertainty stands for the form-factor effects and the latter for neglecting the higher-order corrections.
Note that the value $a_\pi^\text{univ}$ covers a whole interval of numerical values suggested by various theoretical predictions and experiments.
The new result for $R$ represents a significant improvement (by two orders of magnitude) to the current PDG-based value $R^\text{PDG}=11.88(35)\times10^{-3}$.
It should thus be used in future theoretical predictions or experimental analyses.
Since the sum of all the branching ratios of the neutral-pion decays should sum up to 1, $R$ translates into the branching ratios of the two main $\pi^0$ decay modes: $\mathcal{B}(\pi^0\to\gamma\gamma)=98.8131(6)\,\%$ and $\mathcal{B}(\pi^0\to e^+e^-\gamma(\gamma))=1.1836(6)\,\%$~\cite{Husek:2018qdx}.

Such a precise determination of $R$ and related quantities is possible due to the fact that the form-factor normalization $\mathcal{F}_{\pi^0\gamma^*\gamma^*}(0,0)$ drops out in the ratio.
Consequently, solely the shape of $\mathcal{F}_{\pi^0\gamma^*\gamma^*}(0,q^2)$ represents the form-factor dependence of $R$.
In the case of the neutral-pion Dalitz decay, the linear expansion of the form-factor shape is a very good approximation, since the transferred momentum squared is considerably limited by kinematics. 
The slope $a_\pi$ then constitutes the only relevant hadronic quantity.
The region $x\approx1$, where the effect of the term $a_\pi x$ would actually matter the most, is strongly suppressed; cf.\ Eq.~(\ref{eq:R}).
This, together with the experimental evidence that the slope of the form factor is numerically small, allows for such a high (20\,\%) uncertainty on $a_\pi^\text{univ}$ without having a significant impact on the precision of $R$.

\section{Dalitz decays of \texorpdfstring{$\eta^{(\prime)}$}{eta\^{}(')} mesons}

Taking the pion case as a starting point, let us shortly point out the subtleties and difficulties one encounters when investigating the radiative corrections for the Dalitz decays of $\eta^{(\prime)}$ mesons.
The mass of the $\eta$ meson lies above the muon-pair-production threshold, and $\eta^\prime$ is even more massive than the lowest-lying resonances $\rho$ and $\omega$.
In this way, the calculation becomes unavoidably sensitive to the width of the broad $\rho$ resonance.
The form-factor effects are thus not negligible and a particular form-factor model needs to be taken into account.

Naive radiative corrections for the $\eta\to e^+e^-\gamma$ process were published in Ref.~\cite{Mikaelian:1972jn}, where only the numerical value of the mass of the decaying pseudoscalar was changed with respect to the earlier work~\cite{Mikaelian:1972yg}.
In Ref.~\cite{Husek:2017vmo} the list of the NLO corrections in the QED sector is completed and the previous approach~\cite{Mikaelian:1972jn} is improved: We take into account muon loops and hadronic corrections as a part of the vacuum-polarization contribution, 1$\gamma$IR contribution at one-loop level (using a vector-meson-dominance-inspired model incorporating the $\eta$-$\eta^{\prime}$ mixing), higher-order final-state-lepton-mass corrections and form-factor effects (using recent dispersive calculations~\cite{Hanhart:2013vba,Hanhart:2016pcd}).
Moreover, we systematically study additional processes including $\eta^\prime$ decays: $\eta\to \mu^+\mu^-\gamma$, $\eta^\prime\to e^+e^-\gamma$ and $\eta^\prime\to \mu^+\mu^-\gamma$.
Tables of values suitable for interpolation and further use can be found together with Ref.~\cite{Husek:2017vmo}.

The resulting corrections are most significant for the $\eta^\prime\to e^+e^-\gamma$ decay.
The form-factor effects are considerable and change the shape of the resonance peaks.
In particular, the height of the $\omega$ peak is significantly influenced by the radiative corrections.
This might be interesting for the extraction of $\omega$ properties or of the $\omega$-$\eta^\prime$ interplay; such information might be deduced from $\eta^\prime\to\omega\gamma\to e^+e^-\gamma$ or from $\eta^\prime\to\omega\gamma\to\pi^+\pi^-\pi^0\gamma$.
The radiative corrections can be expected to be different for these two decay branches.

\section{The Dalitz decay of \texorpdfstring{$\Sigma^0$}{Sigma\^{}0}}

\begin{figure}[!t]
\vspace{-2mm}
\centering
\subfloat[][]{
\includegraphics[width=0.4\columnwidth]{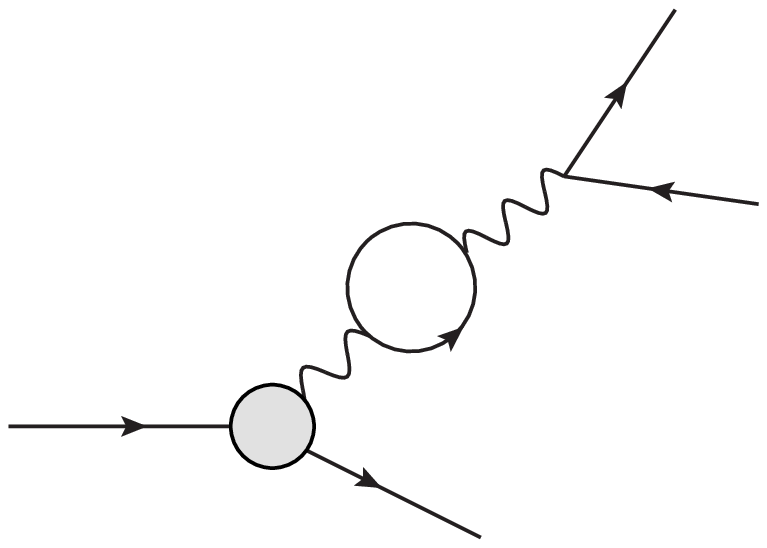}
\label{fig:virta_2}
}
\subfloat[][]{
\includegraphics[width=0.4\columnwidth]{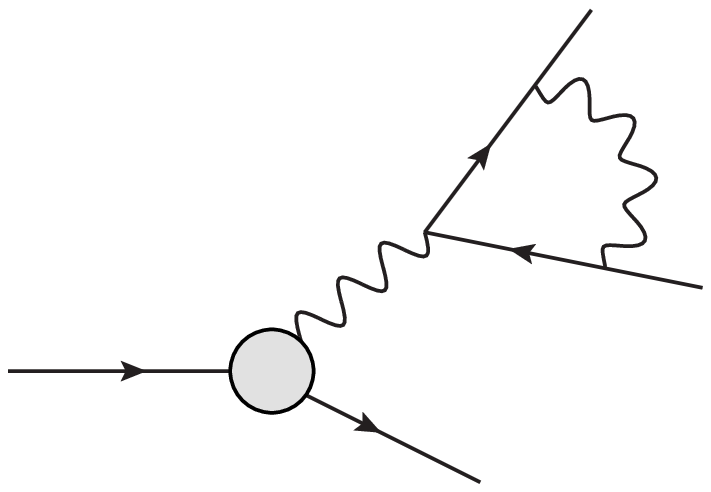}
\label{fig:virtb_2}
}

\subfloat[][]{
\includegraphics[width=0.4\columnwidth]{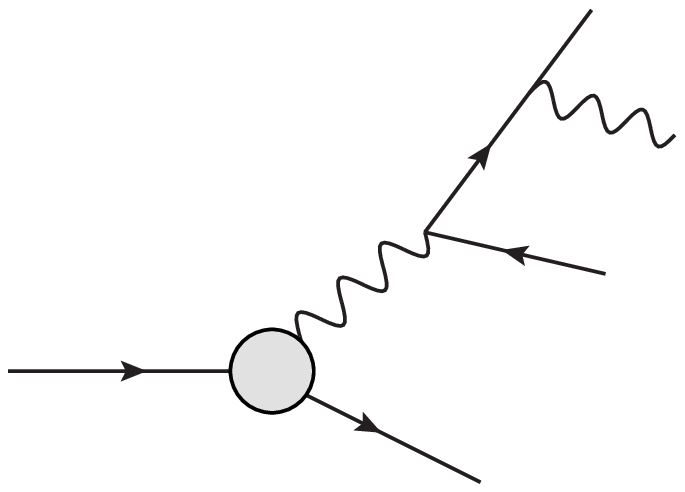}
\label{fig:BS_2}
}
\subfloat[][]{
\includegraphics[width=0.4\columnwidth]{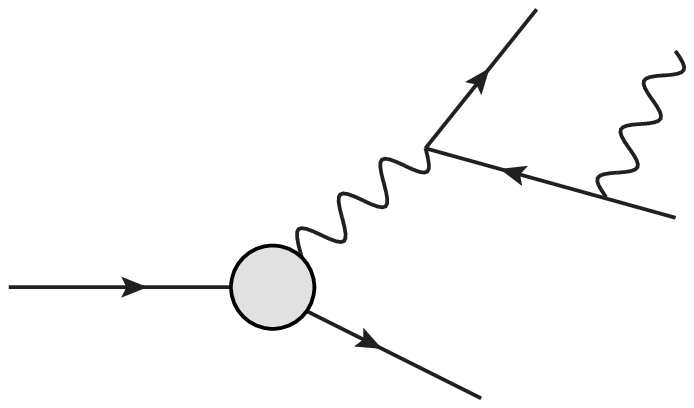}
\label{fig:BS2_2}
}

\subfloat[][]{
\includegraphics[width=0.4\columnwidth]{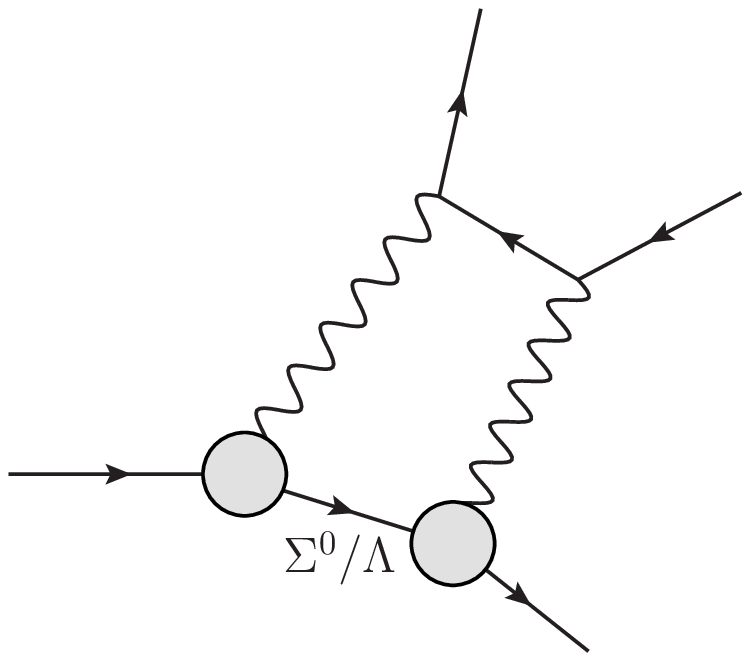}
\label{fig:1gIR_2}
}
\subfloat[][]{
\includegraphics[width=0.4\columnwidth]{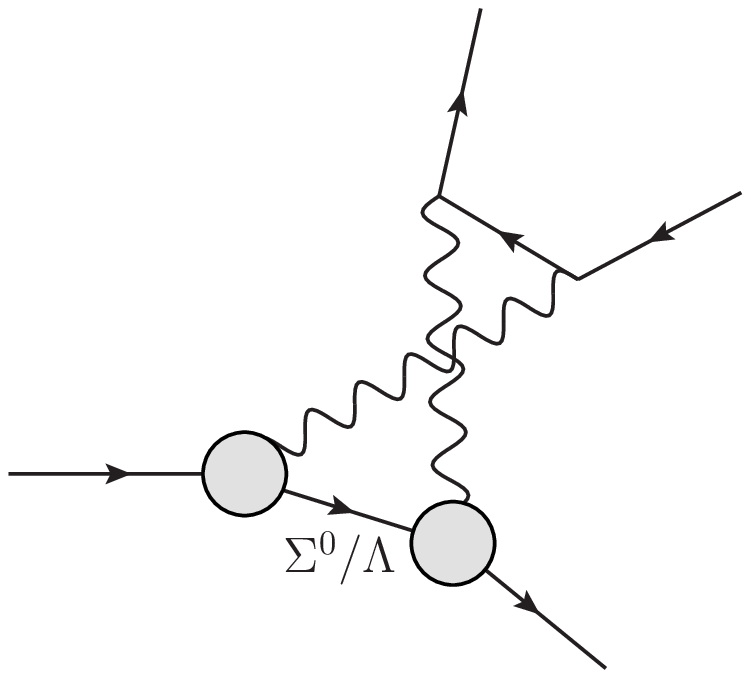}
\label{fig:1gIR2_2}
}

\subfloat[][]{
\includegraphics[width=0.4\columnwidth]{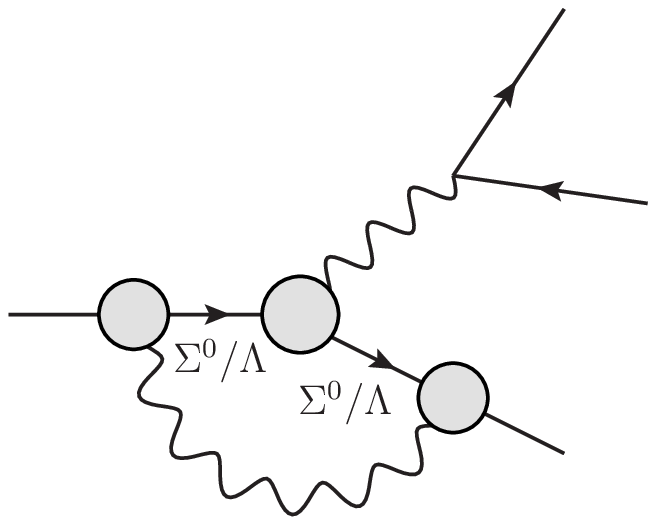}
\label{fig:virt_bar_2}
}
\caption{
\label{fig:baryons}
NLO QED radiative corrections for the decay $\Sigma^0\to\Lambda e^+e^-$: a) lepton-loop vacuum polarization insertion, b) correction to the QED vertex, c) \& d) bremsstrahlung, e) \& f) one-loop one-photon-irreducible contributions, g) $\Sigma^0\Lambda\gamma$ vertex correction.
In the 1$\gamma$IR contribution each diagram comes in two alterations: with $\Sigma^0$ or $\Lambda$ exchanged.
Similarly, there are four diagrams contributing to the transition-form-factor correction g).
}
\end{figure}

So far we discussed radiative corrections to electromagnetic decays in the meson sector.
Possibly, there are related decays in the baryon sector, for which the treatment of radiative corrections might be interesting.

One of the ultimate goals in the low-energy QCD sector is to understand confinement and how are the building blocks distributed within the composite objects---hadrons.
This can be explored using various methods, e.g.\ the electron--nucleon scattering.
We have already seen that some of the structure-related observables---electromagnetic form factors---can be conveniently studied using Dalitz decays.
In the nucleon sector, such a decay does not occur.
However, we can still learn something about the intrinsic structure of nucleons by studying hyperons, which are intimately related to them by replacing the down quark with the strange quark.

In general, performing experiments with hyperons is more complicated since they are very unstable.
The (direct as well as transition) electromagnetic form factors can be accessed at high energies, using electron--positron annihilation into hyperon--antihyperon pair.
At low energies, we would like to make use of a Dalitz decay (of a baryon into another baryon and the electron--positron pair).
Indeed, in this sector there is an interesting decay to be studied in greater detail: $\Sigma^0\to\Lambda e^+e^-$.
Experimentally, in future, it can be investigated with possibly high statistics at FAIR.

The Dalitz decay of $\Sigma^0$ allows us to study the singly-virtual transition form factor within the small window of virtualities up to $M_{\Sigma^0}-M_\Lambda\approx77\,\text{MeV}$.
According to theoretical predictions~\cite{Kubis:2000aa,Granados:2017cib}, the transition radii are very small and of the same order as the NLO QED ratiative corrections.
Consequently, from the experimental point of view, not only the extraction of the form factor parameters will be challenging and require a high-precision measurement, but the good knowledge of the radiative corrections is necessary.

Radiative corrections to the decay rate as well as, within the soft-photon approximation, to the differential decay width were studied in Ref.~\cite{Sidhu:1972rx}.
In our recent work \cite{Husek:2019wmt} we study the inclusive corrections {\em beyond} the soft-photon limit, taking into account the complete set of contributions at NLO (for diagrams see Fig.~\ref{fig:baryons}), i.e.\ including in addition the 1$\gamma$IR contribution (Figs.~\ref{fig:1gIR_2} and \ref{fig:1gIR2_2}) and QED corrections to the $\Sigma^0\Lambda\gamma$ vertex (Fig.~\ref{fig:virt_bar_2}).
By the loop-momenta-power counting, form factors need to be taken into account to regulate the UV behavior of the loop integrals of the two latter mentioned contributions.
For the 1$\gamma$IR contribution, the UV convergence is achieved already for a simple model with constant electric and magnetic form factors: $G_\text{E}(q^2)=0$ and $G_\text{M}(q^2)=\kappa$, where $\kappa$ is related to the magnetic moment.
For the triangle diagram in Fig.~\ref{fig:virt_bar_2}, a stronger UV suppression is necessary.
The effect of these contributions was confirmed to be negligible independently of the chosen model.
Numerically, the main difference between our results~\cite{Husek:2019wmt} and the work~\cite{Sidhu:1972rx} thus lies in the way how the bremsstrahlung contribution is treated.
Regarding this contribution, a low-energy expansion of the form factor can be used and the model dependence can be conveniently scaled out to a large extent in the final correction.

\begin{figure}[!t]
\centering
\includegraphics[width=\columnwidth]{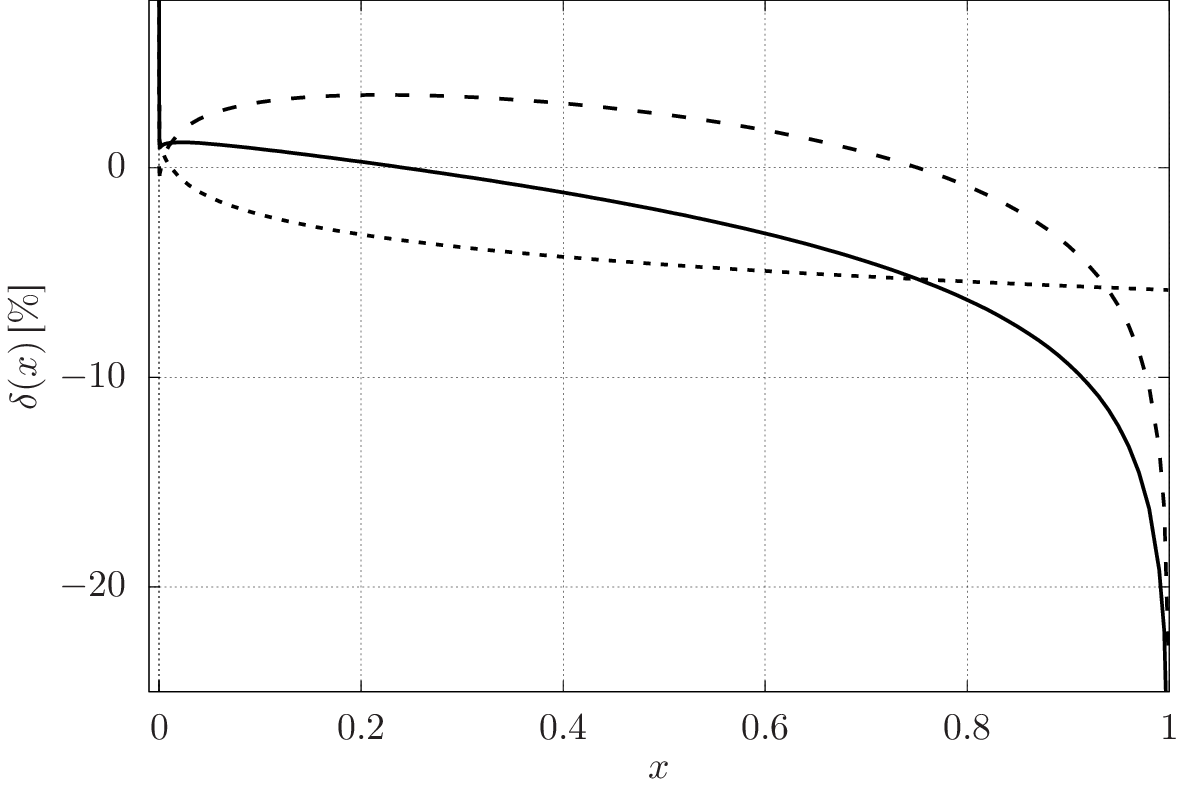}
\caption{
\label{fig:deltax}
The inclusive NLO QED radiative correction $\delta(x)$ (solid line) in comparison to its constituents for the decay $\Sigma^0\to\Lambda e^+e^-$.
The virtual corrections are depicted as the dotted line, the bremsstrahlung as the dashed line.
}
\end{figure}

The radiative corrections to the one-fold differential decay width are shown in Fig.~\ref{fig:deltax}.
Integrating the two-fold differential decay width (taken up to NLO) over the whole Dalitz plot reveals
\begin{equation}
R_{\Sigma_\text{D}^0}
\equiv\frac{\Gamma(\Sigma^0\to\Lambda e^+e^-)}{\Gamma(\Sigma^0\to\Lambda\gamma)}
=5.544(2)\times10^{-3}\,,
\label{eq:R_sigma}
\end{equation}
which is perfectly consistent with the NLO result for the rate given in Ref.~\cite{Sidhu:1972rx} and obtained using a different method:
\begin{equation}
R_{\Sigma_\text{D}^0}^\text{S\&S}
=(5.532+0.627a)\times10^{-3}\,.
\end{equation}
Here, $a$ is related to the magnetic radius; we used $a\equiv\frac16\langle r_\text{M}^2\rangle(M_{\Sigma^0}-M_\Lambda)^2\approx0.0183(26)$~\cite{Kubis:2000aa}.
The result (\ref{eq:R_sigma}) can be again translated into branching ratios: $\mathcal{B}(\Sigma^0\to\Lambda\gamma)=99.4486(5)\,\%$ and $\mathcal{B}(\Sigma^0\to\Lambda e^+e^-)=0.5514(5)\,\%$.

Finally, from Fig.~\ref{fig:deltax} we can estimate the size of the correction to the (magnetic) form-factor slope by taking half of the slope of the curve in the low-$x$ region, however farther from the threshold:
\begin{equation}
\Delta a
\equiv a_\text{(+QED)}-a
\simeq\frac12\frac{\diff\delta(x)}{\diff x}\bigg|_{x=x_0}\,,
\end{equation}
with $\nu^2\ll x_0\ll1$.
We find $\frac12\frac{\diff\delta(x)}{\diff x}\big|_{x=x_0}\approx-3.5\,\%$.
Since this value is bigger than the estimate on $a$ itself, it is indeed essential to consider radiative corrections when extracting the magnetic radius from experiment.

\ \\
The complete sets of NLO radiative corrections in the QED sector for the discussed decays---the Dalitz decays of $\pi^0$~\cite{Husek:2015sma}, $\eta^{\prime}$~\cite{Husek:2017vmo} and $\Sigma^0$~\cite{Husek:2019wmt}---are available and their use in future experimental analyses should be essential.
Finally, note that in Refs.~\cite{Husek:2015sma,Husek:2017vmo,Husek:2019wmt} we study fully inclusive radiative corrections: No momentum or angular cuts on additional photon(s) are applied.

\section*{Acknowledgments}

The presented work is based on collaboration with K.~Kampf, J.~Novotn\'y, S.~Leupold and E.~Goudzovski.

This work has been supported in part by
Grants No.\ FPA2017-84445-P and
SEV-2014-0398 (AEI/ERDF, EU)
and by PROMETEO/2017/053 (GV).


\providecommand{\href}[2]{#2}

\end{document}